\documentclass{emulateapj}

\shorttitle{A forming cluster of galaxies at z=1.6}
\shortauthors{Castellano et al.}

\begin{document}


\title{A photometrically detected forming cluster of galaxies at redshift 1.6 in the GOODS field} 

\author{M. Castellano\altaffilmark{1}, S. Salimbeni\altaffilmark{2}, D. Trevese\altaffilmark{1}, A. Grazian\altaffilmark{2}, L. Pentericci\altaffilmark{2}, F. Fiore\altaffilmark{2}, A. Fontana\altaffilmark{2}, E. Giallongo\altaffilmark{2}, P. Santini\altaffilmark{2}, S. Cristiani\altaffilmark{3}, M. Nonino\altaffilmark{3} and E. Vanzella\altaffilmark{3}}

\altaffiltext{1}{Dipartimento di Fisica, Universit\'{a} di Roma ``La Sapienza'', P.le A. Moro 2, 00185 Roma, Italy}

\email{marco.castellano@uniroma1.it}

\altaffiltext{2}{INAF - Osservatorio Astronomico di Roma, Via di Frascati 33, 00040 Monte Porzio Catone, Italy}
\altaffiltext{3}{INAF - Osservatorio Astronomico di Trieste, Via G.B. Tiepolo 11, 34131 Trieste, Italy}




\begin{abstract}
We report the discovery of a localized overdensity at z$\sim$1.6 in the GOODS-South Field, 
presumably a poor cluster in the process of formation. The three-dimensional galaxy density has been estimated on the basis of well calibrated photometric redshifts from the multiband
photometric GOODS-MUSIC catalog using the (2+1)D technique.
The density peak is embedded in the larger scale overdensity of galaxies  
known to exist at z=1.61 in the area.
The properties of the member galaxies are compared to those of the surrounding field and we found that the two populations are significantly different supporting the reality of the structure. The reddest galaxies, once evolved according to their best fit models, have colors consistent with the red sequence of lower redshift clusters.
The estimated $M_{200}$ total mass of the cluster is in the range $ 1.3 \times 10^{14} - 5.7\times 10^{14} \ M_{\odot}$, depending on the assumed bias factor \textit{b}. An upper limit for the 2-10 keV X-ray luminosity, based on the 1Ms Chandra observations, is $L_X=0.5\cdot 10^{43} \ erg \ s^{-1}$, suggesting that the cluster has not yet reached the virial equilibrium.
\end{abstract}




\section{Introduction}
Clusters of galaxies are the largest and most massive gravitationally bound
structures in the Universe and therefore are excellent laboratories to investigate the
formation and evolution of galaxies and their  interaction
with environment \citep{mil04,nak05,tra05}. For this reason finding
clusters at  high redshift  is the key to understand the origin of the peculiar properties
observed at low redshift, like morphology/color segregation \citep{dress1,dress2} and the presence of the `red
sequence' in color-magnitude diagrams \citep{koda98}.
Studies of X-ray detected massive clusters up to redshifts  $z \sim 1.4$ \citep{toz03,mau04,ros04,mul05,stanf06} 
have found little evolution in their properties with respect to present-day structures despite the large
lookback times ($\sim$ 65\% of the age of the universe). However only very massive clusters have been detected so far, due to the strong dependence of the X-ray luminosity on the gas mass \citep{arnaud}. Alternatively high-z cluster candidates can be selected on the basis of infrared imaging, as in the Spitzer/IRAC Shallow Survey \citep{stanf05}, as long as these structures are dominated by massive elliptical galaxies, whose emission is peaked at 1-2 $\mu m$. Finally surveys based on the Sunyaev-Zeldovich (SZ) effect that is, in principle, redshift independent, do not reach yet the required sensitivity to detect any of the known clusters at $z > 1$ \citep{carls02}.
A completely different approach consists in searching for
concentrations of Ly$\alpha$ emitters near radio galaxies, assumed to be tracers of high density
regions \citep[e.g.,][]{venem07}. This method however, besides selecting only  proto-clusters associated with these peculiar objects, is limited at z $>$ 2 for ground based observations.

Methods used at low redshift become impractical for finding distant clusters. The Friends of Friends (FOF) method \citep{huchra82} can be applied to spectroscopic surveys which have much brighter magnitude limits with respect to photometric ones.
Two-dimensional (2D) photometric surveys,  
which  are  deeper, have been analyzed but
additional  assumptions have to be applied, e.g. on  the galaxy  luminosity
function (LF), as in the Matched Filter algorithm \citep{post96}, or assuming {\it a  priori} the presence
of a  red sequence \citep{gladd05}.
These methods present some difficulties in the redshift range
$1<z<2$, where according to the relevant models and observations we
expect to find the red sequence formation and the first hints of
color segregation \citep{cucciati06}.
To exploit the advantages of deep photometric surveys without any assumption on cluster properties we use a different approach, the (2+1)D algorithm described in \citet{tre07} and in section 3. It is based on an estimate of three dimensional densities  using simultaneously angular positions and photometric redshifts, obtained from multi-band photometry at the deepest observational limits. Although the uncertainty of photometric redshifts is higher than that of spectroscopic ones, they have been already used for the detection of large scale structures \citep{tre07,scoville}.\\
All the magnitude used in the present paper are in the AB system. We adopt a $\Omega_{\Lambda}$=0.7,  $\Omega_{M}$=0.3 and $H_0$=70 Km s$^{-1}$ Mpc$^{-1}$ cosmology.

\section{The GOODS-MUSIC sample}

We use the multicolor GOODS-MUSIC catalog extracted from a deep and wide survey
over the Chandra Deep Field South. The data comprise 14 bands photometry,
including the $U$-band data from the 2.2ESO ($U_{35}$ and $U_{38}$)
and VLT-VIMOS ($U_{VIMOS}$), the ACS-HST data in four bands ($F435W$, $F606W$, $F775W$ and $F850LP$), VLT-ISAAC data
($J$,$H$ and $Ks$) and  Spitzer data provided by IRAC instrument (3.5, 4.5, 5.8 and 8 $\mu m$).
All the images cover an area of 143.2 $arcmin^2$, except for
the U-VIMOS image (90.2 $arcmin^2$) and the H image (78.0
$arcmin^2$). The multicolor catalog contains 14847 extragalactic 
objects, selected either in the 
$z_{850}$ or in the $Ks$ band. We use the $z_{850}$-selected sample ($z_{850} \sim 26$) which contains 9862 galaxies (after excluding AGNs 
and galactic stars). Since the S/N ratio varies along the image a single magnitude 
limit cannot be defined, so the sample is divided in 6 sub-catalogs each with 
a well defined area and magnitude limit in the range 24.65-26.18  \citep[for more details see][]{grazian}.
About 10\% of the galaxies in the sample have spectroscopic redshift. For the other galaxies we have computed photometric redshift using a standard 
$\chi^2$ minimization over a large set of templates obtained from synthetic
spectral models, using the P\'{E}GASE 2.0 libraries \citep{pegase}. See also \citet{giallongo98} and \citet{fontana00} for further discussion of the method.
The comparison with the spectroscopic subsample shows that the accuracy of the photometric redshift is very good, with $\left<|\Delta z/(1+z)|\right> =0.03$ up to  redshift $z=2$. A more detailed description of this catalog can be found in \citet{grazian}.





\section{Finding Algorithm and Cluster Detection}

At low redshift, where background/foreground contaminations are acceptable, the analysis of the surface density is sufficient to detect galaxy clusters \citep[e.g.,][]{abell}.
Similarly, using spectroscopic redshifts it is possible to compute the three-dimensional distance $D_n$ to the n-th nearest neighbor and derive a volume density.

In deep photometric surveys, where spectroscopic redshifts are not available for most of the galaxies, we have developed a method that 
consists in combining, in the most effective way, the   
angular position with the distance computed from the photometric
redshift of each object. The method is described in detail in \citet{tre07} and has been adapted here 
to analyze a composite sample of galaxies from sub-catalogs with different magnitude limits. 
We divide the survey volume in cells whose extension in  different directions 
($\Delta\alpha, \Delta \delta, \Delta z$)
depends on the relevant positional accuracy and thus are elongated in the radial direction. We have chosen the cell sizes small enough to keep an acceptable spatial resolution, while avoiding a useless increase of the computing time.
We have adopted, at $z \sim 1$, $\sim 60 $ Mpc (comoving) in the radial direction and $\sim$ 2.4 arcsec $\sim 40$ kpc (comoving) in transverse direction ($\alpha,\ \delta$). 
For each  cell in space  we then count neighboring objects 
at increasing distance, until a number $n$ of objects is reached. We then assign to the cell a comoving density $\rho = n/ V_n$ where $V_n$ is the comoving volume which
includes the n-nearest neighbors. 
The choice of $n$ is a trade off between spatial resolution and signal-to-noise ratio. For the GOODS field we have fixed n = 15. Note that the algorithm does not use a fixed smoothing lenght, but the resolution increases with density.
\begin{figure}[!ht]
 \centering
\resizebox{\hsize}{!}{
 \includegraphics{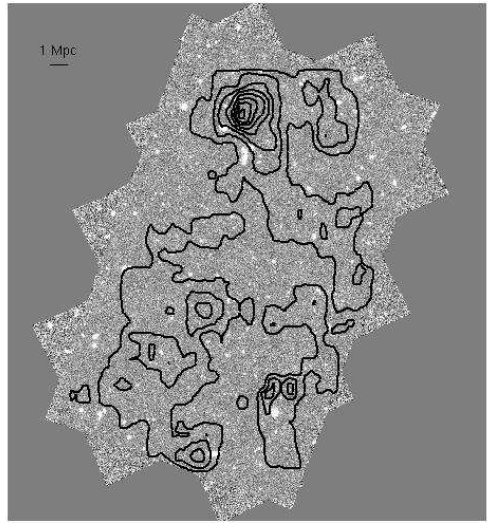}}
\caption{Density isosurfaces at z$\sim 1.6$ (average, average+1$\sigma$ to average+6$\sigma$) superimposed on the ACS $z_{850}$ band image of the GOODS-SOUTH field. The angular dimension of 1 Mpc (comoving) at z=1.6 is indicated in the top-left corner.}
\end{figure}
\begin{figure}[!ht]
 \centering
 \resizebox{\hsize}{!}{\includegraphics{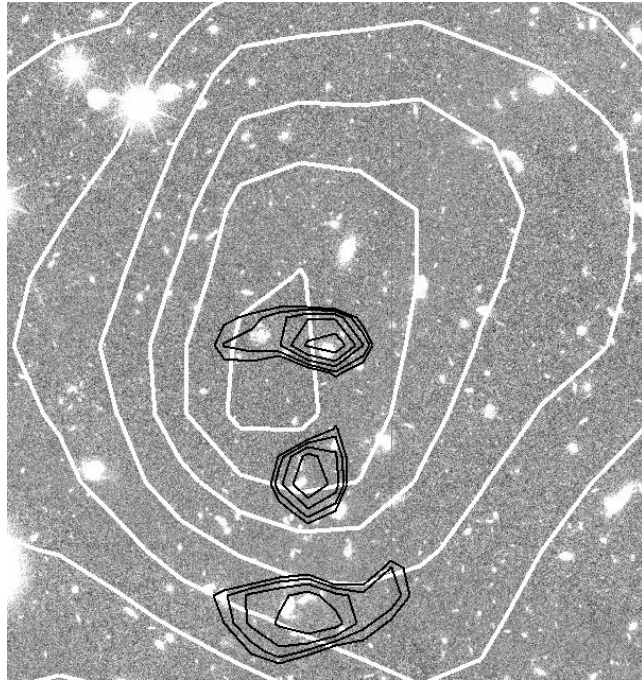}}
\caption{The enlargement of the ACS $z_{850}$ image showing the density peak. White contours are as in Figure 1. Black lines are X-ray contours in the 0.4-3 keV interval. }
\end{figure}
We take into account the increase of limiting absolute magnitude with increasing redshift for a given apparent magnitude limit, assigning a weight $w(z)=1/s(z)$ to each detected galaxy at redshift $z$, where $s(z)$ is the fraction of objects detected with respect to a reference redshift $z_c$ below which we detect all objects brighter
than the relevant $M_c \equiv M_{lim}(z_c)$:
\begin{equation}
s(z)=\frac{\int^{M_{lim(z)}}_{-\infty}\Phi(M)dM}{\int^{M_c}_{-\infty}\Phi(M)dM} 
\end{equation}
$\Phi(M)$ is the redshift dependent galaxy luminosity function computed on the same GOODS-MUSIC catalog (Salimbeni et al. 2007, submitted), $m_{lim}$ is 
the  apparent magnitude limit which, in the GOODS-MUSIC sample, depends on the position,  $M_{lim}(z)$ is 
the corresponding absolute magnitude limit at the given redshift $z$. We use this {\it magnitude correction} to obtain 
a density scale independent of redshift, at least to a first approximation. 
In computing $M_{lim}(z)$ we use k - and evolutionary - corrections for each object computed with the same best fit SEDs used to derive the photometric redshift. 
Galaxy overdensities, which can be regular clusters or groups as well as 
filamentary or diffuse structures, are defined as connected 3-dimensional 
regions with density exceeding
a fixed threshold (see below).
We have computed the average and r.m.s. dispersion of the density field on the comoving volume in the range $0.4<$z$<2.0$. Below the lower redshift limit, the comoving volume sampled is too small to provide a reasonable statistics; on the other hand at redshift above 2 the limiting magnitude correction discussed above becomes less reliable.
We use a density threshold of 2$\sigma$ above the average to identify structures.
Several major localized peaks at different redshifts emerge from the entire field,
some of them already discussed in the literature, e.g. \citet{gilli03}, \citet{tre07}, \citet{diaz}. In this letter we focus on a peak centered at z$\sim$1.6, which is the most significant at z$>$1 in this field. 
The existence of a diffuse structure at this redshift is already known from spectroscopic observations \citep{cimatti02,gilli03,vanzella05a,vanzella05b}, and it is indicated by the presence of 28 objects, distributed over the entire GOODS field, with  redshift in a small interval around z=1.61. We also detect this large scale overdensity at this redshift, diffuse on the entire GOODS field. Within this extended structure we isolate a compact, higher density peak, approximately centered at RA=$03^h 32^m 29.28^s$, DEC=$-27^{\circ} 42' 35.99''$ as shown in Figure 1, that we identify as a cluster, having a total extension of approximately $3 \times 3$ Mpc (comoving). Indeed it is apparently a symmetric structure that, like most lower redshift clusters, is embedded in a wider wall-like or filamentary overdensity \citep{koda05}.
A visual inspection of the ACS images  shows that the structure is centered on a large, star-forming and remarkably irregular galaxy (see Figure 2).

The existence of the structure is confirmed by a comparison of galaxies in the higher density region around the peak with `field' galaxies at the same redshift in the rest of the GOODS field.
The first sample includes 45 galaxies having an associated density of at least $\rho$ = 0.022 $Mpc^{-3}=\bar{\rho}+2\sigma_{\rho}$ and photometric redshift in the range $1.45<$z$<1.75$ (i.e. 1.6 $\pm2\sigma_z$ at that redshift). 
`Field' galaxies are those objects, in the same redshift interval, with associated $\rho<0.022 \ Mpc^{-3}$ (to avoid other smaller overdensities that are present in the field) and outside a square whose side is 2$\times R_A$ (the Abell radius, 2.14 Mpc, comoving, with h=0.7) centered on the overdensity peak, to avoid any contamintion from members of the cluster.
\begin{figure}[!ht]
 \centering
 \resizebox{\hsize}{!}{\includegraphics{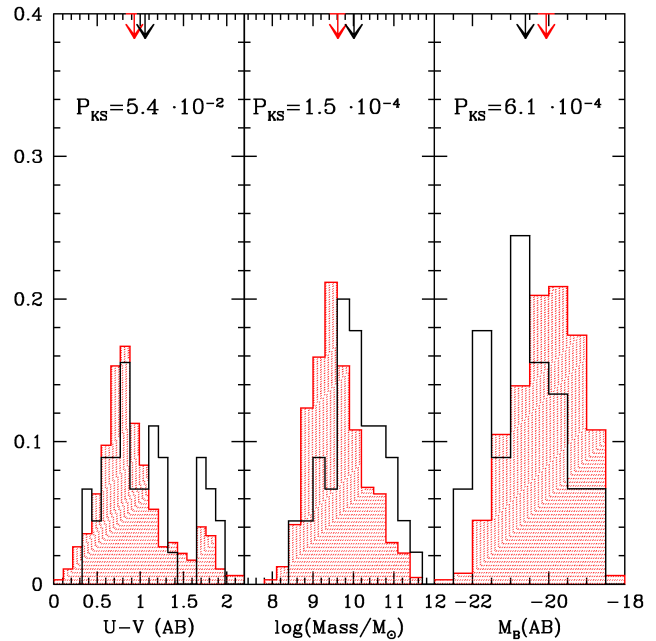}}
\caption{Histograms of the (U-V) color, total stellar mass (in solar units) and $M_B$ (AB) magnitude
for objects selected in the field (shaded histogram) or in the cluster, as described in the text. In each panel we indicate the Kolmogorv-Smirnov probability of the null hypothesis that the two samples are drawn from the same distribution. The averages of the distributions are indicated with arrows.}
\end{figure}
\begin{figure}[!ht]
 \centering
 \resizebox{\hsize}{!}{\includegraphics{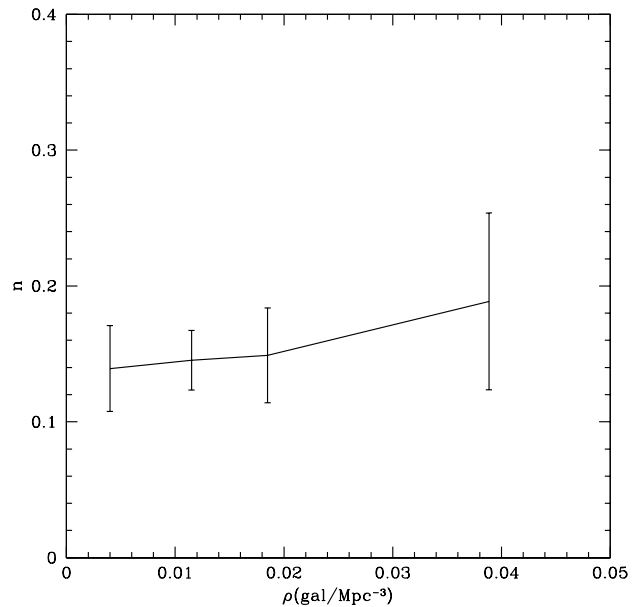}}
\caption{Fraction of red galaxies, selected with (U-V) color, as described in the text, on the entire GOODS field in the redshift interval $1.45<z<1.75$, as a function of the associated density.}
\end{figure}
For these two samples we have computed the distributions of rest frame U-V colors, total stellar mass and rest frame $M_B$ magnitude.  In Figure 3 we plot the resulting distributions. We find that the galaxies in the overdensity are significantly brighter and more massive than the field galaxies  and there is also evidence of a color segregation, although the overdensity is mainly formed by blue star-forming objects with only few red and passively evolving galaxies. According to a Kolmogorov-Smirnov test, the probability of the null hypothesis that the mass, $M_B$ magnitude and (U-V) color distributions inside and outside the overdensity are randomly drawn from the  same distribution is $1.5 \cdot 10^{-4}$, $6.1 \cdot 10^{-4}$ and $5.4 \cdot 10^{-2}$ respectively. Therefore, at least for stellar mass and magnitudes, we can conclude that the two populations are significantly different, thus supporting the evidence that this is a real structure.
The (U-V) color distribution in the overdensity is slightly different from that of the surrounding field. This is also shown by the fraction of red galaxies as a function of the environmental density.  Red galaxies are defined as those with $(U-V)> -0.08 \cdot (M_B + 20) + 1.36$, the reference color corresponding to the locus of the minimum between the blue and red populations at this redshift in the GOODS-MUSIC catalog, as derived by Salimbeni et al. 2007 (submitted). In Figure 4 we show the fraction of red galaxies as a function of the environmental density on the entire GOODS field. The highest density bin corresponds, in practice, to the cluster. 
The color segregation is much less evident than at lower redshift \citep{tre07}, to the point that, given the large uncertainties, it is not possible to exclude the absence of any trend of the red fraction with density. We have performed a fit of the data in the two different hypothesis of (i) a constant value of the red fraction; (ii) linear growth of the red fraction with density. On the basis of a two-tails $\chi^{2}$ test, both solutions are compatible with the data ($P(\chi^{2})>15\%$).

We also find, in agreement with \citet{cucciati06} that, at this redshift, the fraction of `red' galaxies is much lower than the fraction of the `blue' ones, for every value of the environmental density. This implies that, in this case of a small forming structure, the detection of the overdensity through its reddest members would have been more difficult.

\section{Properties of the Structure}
A more detailed analysis of the structure can be carried out from the study of the color-magnitude diagram.
In the upper panel of Figure 5 we show the observed ($z_{850}$-$K_s$) vs. $K_s$ color-magnitude diagram for galaxies located within the overdensity. 
The best-fit models of  the 9 reddest galaxies ($z_{850}-K_s> 2.2$) , from the library of \citet{bruz03}, show that seven are fit by passively evolving models with a short star-formation e-folding time ($\tau$ $\sim$ $0.3$ Gyr), ages larger than 2 Gyr (corresponding to $z_{form}>3$), total stellar mass around $10^{11} \ M_{\odot}$ and no significant dust reddening. 
Of the remaining two objects, one is similar to the other seven but is fit by a model with $\tau$ $\sim$ $1$ Gyr and a moderate star-formation rate. The last object, which is the faintest of the nine galaxies in the $K_s$ and $z_{850}$ bands, is fit by a young (age $<$ 100 Myr) and extremely star forming galaxy, reddened by dust (E(B-V) $\sim$ 0.9).
According to the best-fit models, we have evolved these galaxy populations from z=1.6 to z=1.24 to compare them to the `red sequence' found for the cluster RDCS J1252.9-2927 at z=1.24 \citep{blake,lidman,demarco} and represented by the lines in the lower panels of Figure 5. 
We find a good agreement with the `red sequence' obtained by these authors. It is remarkable that at decreasing redshift the objects become less dispersed around the linear fit obtained for lower z clusters.
We note that, of this nine red galaxies, only the dusty, high-SFR object remains significantly away from the red sequence, while the object with $\tau$ $\sim$ $1$ Gyr appears to be exactly in the linear sequence in one of the two plots (($i_{775}-z_{850}$) vs $z_{850}$) and at some distance in the other (($z_{850}$-$K_s$) vs. $K_s$).
\begin{figure}[!ht]
\centering
\resizebox{\hsize}{!}{\includegraphics{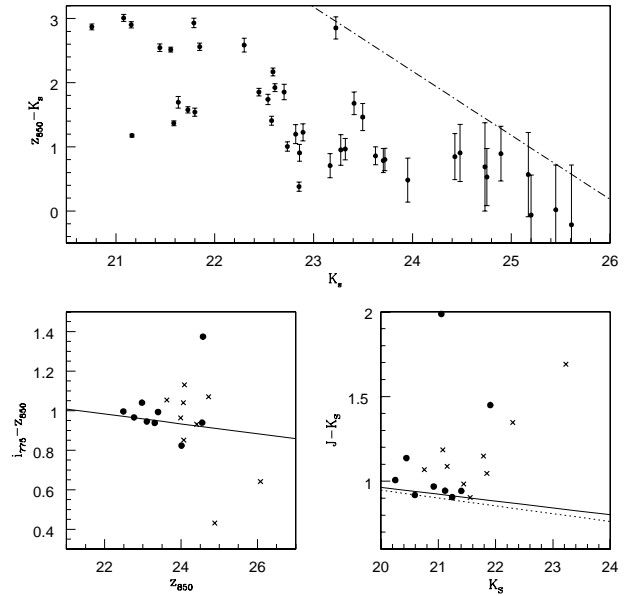}}
\caption{Upper panel: color magnitude diagram ($z_{850}$-$K_s$) vs. $K_s$ for the galaxies of the sample, the dashed line corresponds to the observed magnitude limit $z_{850}$=26.18. Lower panels: color-magnitude diagrams ($i_{775}-z_{850}$) vs $z_{850}$ (left) and ($J$-$K_s$) vs $K_s$ (right) with observed colors (crosses) and colors evolved, as described in the main body, up to redshift 1.24 for the same objects (filled dots). The continuous lines are the linear fits of the `red sequence' in the cluster RDCS J1252.9-2927 at z=1.24, by \citet{blake} (left panel) or by \citet{demarco} (right panel). The dotted line in the right panel is the fit by \citet{lidman} for the same cluster.}
\end{figure}

We have estimated the mass associated with this overdensity using the galaxy density contrast and adapting to our case the method used for spectroscopic data at higher z by \citet{steidel}:
$$ M=\bar{\rho_{u}} \cdot V \cdot (1+\delta_{gal}/b), $$
where  $\bar{\rho_{u}}$ is the average density of the Universe and $b$ is the bias factor that we assume in the interval $1<b<3$ \citep{arno}.
First we count the objects in the photometric redshift range $1.45<$ z $<1.75$ as a function of the cluster-centric radius. Then we perform a statistical subtraction of the background/foreground field galaxies in each magnitude bin computing their mean surface density in the rest of the field and in the same redshift interval. Finally the 3d galaxy density contrast $\delta_{gal}$ is computed assuming spherical symmetry of the structure.
The mass $M_{200}$, inside the radius within which the density contrast is  $\delta_{tot}=\delta_{gal}/b$ $\sim$ 200 \citep{carlberg}, is in the interval $ 1.3 \times 10^{14} - 5.7\times 10^{14} \ M_{\odot}$ depending on the assumed bias factor.
When the value of the radius is equal to $R_A$, as defined in section 2, we obtain a rough estimate of the richness and mass following the Abell classification. After the statistical subtraction of the field objects, the number of galaxies within the Abell radius is $N$ = 41, of which 28 are between $m_3$ and $m_3+2$, comparable to that of a richness class 0 cluster. The corresponding mass is in the range $1.4 \times 10^{14}-4.1 \times 10^{14} \ M_{\odot}$, indeed typical for a poor cluster \citep{sar}.
However, these richness and mass have to be considered as lower limits, because the density contrast is computed with respect to the average density of the GOODS field, which, in the range $1.45< z_{phot} <1.75$, is dominated by the large scale structure discussed before that comprise the cluster itself.

Given that the area was covered by deep Chandra observations \citep{giacco}, we have checked if the cluster has associated X-ray emission, which could be an indication of the virialization status.
An inspection of the 1Ms Chandra observation of the field shows that X-ray emission is detected within the core and it is divided in three different clumps (Fig. 2): in particular the elongated X-ray clump located at the center of the structure includes several overdensity members, altough it could in part be contaminated by a Seyfert galaxy at z$\sim 0.6$. 
The total count rate in the interval 0.3-4 keV is $(7.40\pm 1.65) \times 10^{-5}$.
Assuming a thermal spectrum with T=3 keV and Z=0.2 Z $_{\odot}$, we obtain a total flux $5\cdot 10^{-16}$erg $s^{-1}$ $cm^{-2}$  (in the interval 0.3-4 keV) and a total emission $L_X \sim 0.5 \cdot 10^{43} \ erg \ s^{-1}$ (in the interval 2-10 keV). However, the uncertainty on the flux is of a factor 2 depending on the assumed temperature of the thermal spectrum. This total flux is lower than what is expected for a cluster of this mass and richness \citep[$L_X \gtrsim 10^{43} \ erg \ s^{-1}$, ][]{ros2000,ledlow}. The low luminosity and the irregular morphology of the X-ray emission thus indicate that the group/poor cluster has not yet reached its virial equilibrium.

\section{Summary and Conclusions}

We have used an estimation of the three dimensional density of galaxies, through the (2+1)D algorithm, to detect structures in the GOODS-SOUTH field. The one at z$\sim 1.6$ presented in this paper is the most significant at $z>1$ in the entire GOODS field, and it consists of a density peak embedded in an already known large scale structure.
The reality of this structure is confirmed by the fact that the mass and luminosity distributions of its galaxy population are significantly different from  that of the surrounding field at the same redshift. The overdensity consists mainly of star-forming galaxies and few passively evolving objects.
The reddest galaxy members in the overdensity are consistent with being the progenitors of the red sequence galaxies known at lower redshifts. 
We have estimated the richness of the cluster (richness class 0 in the Abell classification) and its $M_{200}$ mass, that is in the range $1.3 \times 10^{14}-5.7 \times 10^{14} \ M_{\odot}$. The X-ray data from the Chandra 1 Ms observation show the presence of faint, clumpy emission within the core. This irregular morphology and the low total luminosity indicates that the structure has not yet reached its virial equilibrium.
This also implies that the detection of a cluster of this kind, through X-ray observations, would have been unfeasible. From the complete photometric and X-ray analysis we can conclude that this structure is probably a forming cluster of galaxies.
\textit{A posteriori}, we can say that the detection through other techniques, such as those employed by the `red sequence surveys' \citep{gladd05} or clustering of IRAC sources \citep{stanf05}, would have been more difficult because most of the cluster members are blue star-forming objects and there is only a small number of galaxies in the red-sequence. In addition, the absence of any peculiar/active source is not surprising given that these kind of objects trace only a small fraction of the high-z overdensities, as discussed in \citet{venem07}. In conclusion, the use of photometric redshifts made possible the individuation of a structure not easily detectable otherways.
However, spectroscopic observations of the overdensity are required to have a precise knowledge of the real members of the cluster and measure its velocity dispersion. Nonetheless, we can state that this structure shows the properties expected for a forming cluster, with hints of color and mass segregation, and that it will probably end up in a moderately rich and virialized present-day cluster.

\acknowledgments
We thank Dr. Michael Strauss for his valuable suggestions.
We thank the whole GOODS Team for providing all the imaging
material available worldwide. Observations have been carried out using
the VLT at the ESO Paranal Observatory under Program
IDs LP168.A-0485 and ID 170.A-0788 and the ESO Science Archive under
Program IDs 64.O-0643, 66.A-0572, 68.A-0544, 164.O-0561, 163.N-0210
and 60.A-9120.

\clearpage




\clearpage


\begin{thebibliography}{}
\bibitem[Abell(1958)]{abell}Abell, G. O. 1958, ApJS 3, 211
\bibitem[Arnaud et al.(2005)]{arnaud}Arnaud, M., Pointecouteau, E., \& Pratt, G. W.  2005, A\&A, 441, 893
\bibitem[Arnouts et al.(1999)]{arno}Arnouts, S., Cristiani, S., Moscardini, L., et al., 1999, MNRAS, 310, 540
\bibitem[Blakeslee et al.(2003)]{blake} Blakeslee, J. P., Franx, M., Postman, M., et al. 2003, ApJ, 596L, 143 
\bibitem[Bruzual \& Charlot(2003)]{bruz03}Bruzual, G. \& Charlot, S., 2003, MNRAS 344, 1000
\bibitem[Carlberg et al.(1997)]{carlberg} Carlberg, R. G., Yee, H. K. C., Ellingson, E., \ 1997, ApJ, 478, 462
\bibitem[Carlstrom et al.(2002)]{carls02} Carlstrom, J.~E.,
Holder, G.~P., \& Reese, E.~D.\ 2002, \araa, 40, 643
\bibitem[Cimatti et al.(2002)]{cimatti02} Cimatti, A., Mignoli, M., Daddi, E., et al. 2002, A\&A, 392, 395
\bibitem[Cucciati et al.(2006)]{cucciati06} Cucciati, O., Iovino, A., Marinoni, C., et al., 2006, A\&A, 458, 39 
\bibitem[Demarco et al.(2007)]{demarco}Demarco, R.,  Rosati, P., Lidman, C., et al. 2007, astro-ph/0703153
\bibitem[D\'{\i}az S\'{a}nchez et al.(2007)]{diaz} D{\'{\i}}az-S{\'a}nchez, A., Villo-P{\'e}rez, I., P{\'e}rez-Garrido, A. \& Rebolo, R, 2007, MNRAS, tmp, 252
\bibitem[Dressler(1980)]{dress1}Dressler, A.  1980, AJ 236, 351
\bibitem[Dressler et al.(1997)]{dress2}Dressler, A., Oemler, A. Jr., Couch, W. J. et al., 1997, ApJ, 490, 577
\bibitem[Fioc \& Rocca-Volmerange(1997)]{pegase}Fioc, M., \& Rocca-Volmerange, B., 1997, A\&A, 326, 950
\bibitem[{{Fontana} et~al.(2000){Fontana}, {D'Odorico}, {Poli}
  et~al.}]{fontana00}
{Fontana} A., {D'Odorico} S., {Poli} F., et~al., Nov. 2000, \aj, 120, 2206
\bibitem[Giacconi et al.(2002)]{giacco}Giacconi, R., Zirm, A., \& Wang, J.  2002, ApJS, 139, 369
\bibitem[{{Giallongo} et~al.(1998){Giallongo}, {D'Odorico}, {Fontana}
et~al.}]{giallongo98}{Giallongo} E., {D'Odorico} S., {Fontana} A., et~al., Jun. 1998, \aj, 115, 2169
\bibitem[Gilli et al.(2003)]{gilli03}Gilli, R., Cimatti, A., Daddi, E., et al.  2003, ApJ, 592, 721
\bibitem[Gladders \& Yee(2005)]{gladd05}Gladders, M. D., \& Yee, H. K. C., 2005, ApJS, 157, 1 
\bibitem[{{Grazian} et~al.(2006){Grazian}, {Fontana}, {de Santis}
  et~al.}]{grazian}
{Grazian} A., {Fontana} A., {de Santis} C., et~al., Apr. 2006,
  \aap, 449, 951
\bibitem[Huchra \& Geller(1982)]{huchra82} Huchra, J., \& Geller, M. 
1982, ApJ, 257, 423  
\bibitem[Kodama et al.(1998)]{koda98}Kodama, T., Arimoto, N., Barger, A.J., \& Arag\'on-Salamanca, A. 1998, A\&A, 334, 99
\bibitem[Kodama et al.(2005)]{koda05}Kodama, T., Tanaka, M., Takayuki, T., et al., 2005, PASJ, 57, 309
\bibitem[Ledlow et al.(2003)]{ledlow} Ledlow, M.J., Voges, W., Owen, F.N. \& Burns, J.O., 2003, AJ, 126, 2740
\bibitem[Lidman et al.(2004)]{lidman} Lidman, C., Rosati, P., Demarco, R., et al., 2004, A\&A, 416, 829
\bibitem[{{Maughan} {et~al.}(2004){Maughan}, {Jones}, {Ebeling}, \&
  {Scharf}}]{mau04}
{Maughan}, B.~J., {Jones}, L.~R., {Ebeling}, H., \& {Scharf}, C. 2004, \mnras,
  351, 1193
\bibitem[{{Miley} {et~al.}(2004){Miley}, {Overzier}, {Tsvetanov}, {Bouwens},
  {Ben{\'{\i}}tez}, {Blakeslee}, {Ford}, {Illingworth}, {Postman}, {Rosati},
  {Clampin}, {Hartig}, {Zirm}, {R{\" o}ttgering}, {Venemans}, {Ardila},
  {Bartko}, {Broadhurst}, {Brown}, {Burrows}, {Cheng}, {Cross}, {De Breuck},
  {Feldman}, {Franx}, {Golimowski}, {Gronwall}, {Infante}, {Martel},
  {Menanteau}, {Meurer}, {Sirianni}, {Kimble}, {Krist}, {Sparks}, {Tran},
  {White}, \& {Zheng}}]{mil04}
{Miley}, G.~K., {Overzier}, R.~A., {Tsvetanov}, Z.~I., {et~al.} 2004, \nat,
  427, 47
\bibitem[{{Mullis} {et~al.}(2005){Mullis}, {Rosati}, {Lamer}, {B{\" o}hringer},
  {Schwope}, {Schuecker}, \& {Fassbender}}]{mul05}
{Mullis}, C.~R., {Rosati}, P., {Lamer}, G., {et~al.} 2005, \apjl, 623, L85
\bibitem[{{Nakata} {et~al.}(2005){Nakata}, {Bower}, {Balogh}, \&
  {Wilman}}]{nak05}
{Nakata}, F., {Bower}, R.~G., {Balogh}, M.~L., \& {Wilman}, D.~J. 2005, \mnras,
  357, 679
\bibitem[Postman et al.(1996)]{post96}Postman, M, Lubin, L. M., Gunn, J. E., et al. 1996, AJ, 111, 615
\bibitem[Rosati et al.(2002)]{ros2000}Rosati, P., Borgani, S., \& Norman, C., 2002, ARA\&A, 40, 539
\bibitem[{{Rosati} {et~al.}(2004){Rosati}, {Tozzi}, {Ettori}, {Mainieri},
  {Demarco}, {Stanford}, {Lidman}, {Nonino}, {Borgani}, {Della Ceca},
  {Eisenhardt}, {Holden}, \& {Norman}}]{ros04}
{Rosati}, P., {Tozzi}, P., {Ettori}, S., {et~al.} 2004, \aj, 127, 230
\bibitem[Sarazin(1988)]{sar}Sarazin, C. L. 1988, X-ray Emission from Clusters of Galaxies
, (Cambridge University Press, Cambridge)
\bibitem[Scoville et al.(2007)]{scoville} Scoville, N., Aussel, H., Benson, A., Blain, A., et al.,  2006, astro-ph/0612384
\bibitem[Stanford et al.(2005)]{stanf05}Stanford, S. A.; Eisenhardt, P. R., Brodwin, M., et al., 2005, ApJ, 634L, 129 
\bibitem[Stanford et al.(2006)]{stanf06}Stanford, S. A., Romer, A. K., Sabirli, K., et al.,  2006, ApJ, 646L, 13 
\bibitem[Steidel et al.(1998)]{steidel}Steidel, C.C., Adelberger, K.L., Dickinson, M., et al., 1998, ApJ 492, 429
\bibitem[{{Tozzi} {et~al.}(2003){Tozzi}, {Rosati}, {Ettori}, {Borgani},
  {Mainieri}, \& {Norman}}]{toz03}
{Tozzi}, P., {Rosati}, P., {Ettori}, S., {et~al.} 2003, \apj, 593, 705
\bibitem[{{Tran} {et~al.}(2005){Tran}, {van Dokkum}, {Illingworth}, {Kelson},
  {Gonzalez}, \& {Franx}}]{tra05}
{Tran}, K.~H., {van Dokkum}, P., {Illingworth}, G.~D., {et~al.} 2005, \apj,
  619, 134
\bibitem[Trevese et al.(2007)]{tre07}Trevese, D., Castellano, M., Fontana, A., \& Giallongo, E., 2007, A\&A 463, 853
\bibitem[Vanzella et al.(2005)]{vanzella05a} Vanzella, E., Cristiani, S., Dickinson, M., Kuntschner, H.,  et al. 2005, A\&A 434, 53
\bibitem[Vanzella et al.(2006)]{vanzella05b} Vanzella, E., Cristiani, S., Dickinson, M., Kuntschner, H.,  et al. 2006, A\&A 454, 423
\bibitem[Venemans et al.(2007)]{venem07} Venemans, B. P., R\"{o}ttgering, H. J. A., Miley, G. K., van Breugel, W. J. M., et al.,  2007, A\&A 461, 823
\end{thebibliography}
\end{document}